\documentclass[a4paper,11pt]{article}
\usepackage{fullpage,amsmath,epsf,amssymb,textcomp,cancel,cite,url}
\usepackage[hang,small]{caption}                                        
\newcommand{\deriv}[2]{\frac{{\rm d} #1}{{\rm d} #2}}                   
\newcommand{\av}[1]{\langle{#1}\rangle}                                 
\newcommand{\abs}[1]{\left|#1\right|}                                   
\newcommand{\eqn}[1]{\begin{equation}{#1}\end{equation}}                
\newcommand{\eqns}[1]{\begin{equation}{\begin{aligned}#1\end{aligned}}\end{equation}}   
\newcommand{\figref}[1]{Fig.~\ref{#1}}                                  
\newcommand{\eqnref}[1]{(\ref{#1})}                                     
\DeclareMathOperator{\erfc}{erfc}                                       

\title{Discontinuous Transition in a Boundary Driven Contact Process}
\author{A. Costa, R. A.\ Blythe and M. R.\ Evans \\[2ex]
\small{SUPA, School of Physics \& Astronomy, University of Edinburgh, Edinburgh EH9 3JZ, UK}}

\begin{document}

\maketitle

\begin{abstract}
The contact process is a stochastic process which exhibits a
continuous, absorbing-state phase transition in the Directed
Percolation (DP) universality class. In this work, we consider a
contact process with a bias in conjunction with an active wall. This
model exhibits waves of activity emanating from the active wall and,
when the system is supercritical, propagating indefinitely as
travelling (Fisher) waves. In the subcritical phase the activity is
localised near the wall. We study the phase transition numerically and
show that certain properties of the system, notably the wave velocity,
are discontinuous across the transition. Using a modified Fisher
equation to model the system we elucidate the mechanism by which the
the discontinuity arises. Furthermore we establish relations between
properties of the travelling wave and DP critical exponents.
\end{abstract}

\section{Introduction}

The contact process \cite{Harris1974} exhibits a nonequilbrium phase
transition from an absorbing state, where the system ends up
in an inactive configuration, to an active, fluctuating
state.  Although originally introduced as a
microscopic model for epidemic spreading, this lattice model and its
relatives have been used to describe a variety of  systems
including percolation, wetting, reaction-diffusion processes,
branching and annihilating random walks \cite{Hinrichsen2000}
and phase transitions in more exotic settings, such
as between two turbulent states in nematic liquid crystals
\cite{Takeuchi2009},
and in proliferating microbial
populations under gravity \cite{Barrett-Freeman2008}.  
The continuous phase
transition from the inactive to active phases falls into the Directed
Percolation (DP) universality class which is thought to pertain for
any such microscopic model exhibiting an absorbing state phase
transition, in the absence of any additional symmetries or
conservation laws \cite{Janssen1981,Grassberger1982}.

The nature of the transition may be modified by the introduction of an
active boundary into the system that ensures that the activity in the
system never dies out
\cite{Chen1999,Froejdh2001,Blythe2001,Henkel2008}.
These studies have mainly focused on
surface critical behaviour and the emergence of new surface critical
exponents.
In the counterpart
to the inactive phase the activity is confined to the boundary region
whereas in the counterpart to the active phase the activity spreads
from the boundary through the system.   
The spreading of activity through the system from the
boundary is conveniently illustrated in the mean-field description of
the DP universality class \cite{Hinrichsen2000} which has the same form as the 
Fisher-KPP\footnote{Here Fisher-KPP
stands for Fisher-Kolmogorov-Petrovskii-Piskunov. It is alternatively
called the \emph{Fisher-Kolmogorov} or \emph{Fisher} equation.}
equation \cite{Saarloos2003}. This nonlinear partial differential equation exhibits
travelling wave solutions in which the active phase invades inactive
regions with a well-defined velocity for the domain wall that separates the two regions.

In this work we consider a variation of the contact process with the
two features of an active boundary and advection away from the
boundary which we refer to as the Driven Asymmetric Contact Process
(DACP).  The Asymmetric Contact Process has been previously studied in
the absence of boundaries and yields a phase transition in the DP
universality class, as expected
\cite{Schonmann1986,Sweet1997,Schinazi1994}.  The addition of an
active boundary, which serves to drive the system, then modifies the
nature of the transition as described above \cite{Blythe2001}.
However, what we discover is that the DP transition now has a
discontinuous aspect in the sense that the velocity of the wavefront
which carries the activity from the boundary jumps discontinuously at
the critical point.  Thus, intriguingly, the continuous DP transition
is accompanied by a velocity discontinuity in the presence of an
active boundary and advection.

There has been recent interest in the addition of an advection term to
the Fisher-KPP equation in the presence of a boundary since a Galilean
transformation no longer serves to remove the advection term (as would
be the case in the absence of boundaries).  When the advection is
directed {\em towards} the boundary there is a competition between the
advective velocity and the Fisher wave velocity, leading to a
phase transition \cite{Barrett-Freeman2008,Derrida2007,Simon2008} from a phase with
activity localised near the boundary to one where the Fisher wave
invades the whole system. The addition of noise into this system leads
to a more complicated scenario where, in the case of  a reflecting boundary studied in
\cite{Barrett-Freeman2010}, the low activity phase could either be localised near the boundary or 
be the absorbing state.
However, the discontinuous transition  observed in that work where the bulk density jumps at the transition, is distinct from that studied in the present work.

To understand the transition we observe, we consider a phenomenological mean-field description of
the system in the form of a  Fisher-KPP equation which
includes the effects of asymmetry in the form of an advective term. 
We show that the solution can be
thought of as a  Fisher wave moving within an envelope given by the
stationary density profile.  Thus, in the subcritical regime the
system exhibits attenuated  waves with non-zero velocity but
whose amplitude decays to zero away from the boundary.  In the
supercritical regime, on the other hand, the stationary profile has a
non-zero limit far from the boundary, thus the Fisher wave propagates
into the bulk with a constant amplitude and non-zero velocity.  This
is the mechanism for the observed discontinuity of the velocity.

The paper is organised as follows: In section 2 we define the
microscopic model and in section 3 we present numerical evidence for the discontinuous
phase transition. In section 4 we discuss the observed scaling behaviour
and identify the Directed Percolation critical exponents.
In section 5 we discuss a mean field theory in the form of a modified Fisher-KPP
equation. Using the mean field picture we then re-examine  in section 6
the simulation results to show that a similar picture holds. We conclude
in section 7.

\section{Microscopic model}

\begin{figure}[tb]                     
        \begin{center}
                \leavevmode             
                \epsfclipon
                \epsfysize=25mm         
                \epsffile{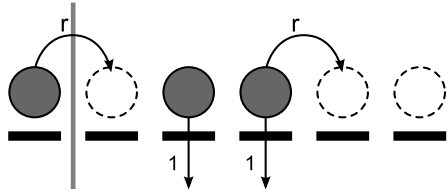}
        \end{center}
    \caption{The Driven Asymmetric Contact Process. In this model, the first site is kept active. Active sites can activate inactive sites to the right of them at rate $r$ or become inactive with  rate~1.}
    \label{fig:DACP}
\end{figure}

The driven asymmetric contact process (hereafter, DACP) is a stochastic model defined
on a one-dimensional lattice.  A microscopic configuration is specified by the set of occupation numbers $\left\{\tau(i) \right\}$ where $\tau_t(i) = 0\,(1)$ indicates that site $i$ is inactive (active) at time $t$. The leftmost site of the system is
kept permanently active, $\tau_t(0)=1 \; \forall t$, making the process
driven. Active sites can activate inactive sites directly to the right
of them at rate $r$ or become inactive with rate 1.
See~\figref{fig:DACP} for an illustration.

\begin{figure}[tb]                     
        \begin{center}
                \leavevmode             
                \epsfclipon
                \epsfxsize=50mm         
                \epsffile{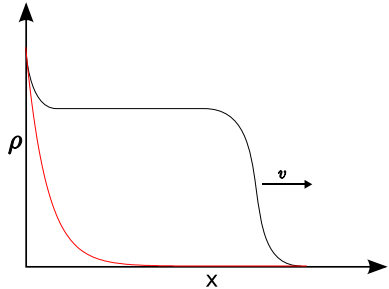}
        \end{center}
    \caption{A sketch of the density profile of the two states. In the subcritical state (red) the density decays exponentially whilst in the supercritical it propagates at a constant bulk density.}
    \label{fig:TwoStates}
\end{figure}

The driving from the left boundary eliminates the absorbing state
associated with the basic contact process. However as the rate $r$
goes through a critical value $r_c$ there is still a continuous phase
transition from a phase where the active sites can only spread a
finite distance from the left boundary to a phase where they can
spread to infinity. See~\figref{fig:TwoStates} for an illustration of
the density profile in the two states.

An interesting and attractive feature of the model is that the total asymmetry of the activation
dynamics means that the state of site $N_1$ is independent from that of $N_2$ for $N_2 > N_1$.
This implies that any calculation or simulation performed on a finite system of $N$ sites gives the \emph{exact} behaviour for the first $N$ sites of an infinite system.  In other words, the introduction of a right boundary does not introduce any finite-size effects. 

We now summarise previous work on asymmetric contact processes. Three
mathematical papers~\cite{Schonmann1986,Sweet1997, Schinazi1994} have
looked at how introducing asymmetry into the contact process (without
a boundary drive) leads to the emergence of a second order parameter,
the probability that the origin is active as $t\rightarrow\infty$, in 
addition to the probability that the process remains active indefinitely. At
total symmetry the two order parameters coincide and at total asymmetry the
second disappears. An approximate analytical and numerical study of
the totally asymmetric contact process (without a boundary
drive)~\cite{Tretyakov1997} found a continuous transition at the
critical rate $r_c = 3.306(4)$ with the critical exponent
$\beta=0.2760(1)$ which is in agreement with the  value for the DP order parameter exponent in one spatial dimension.
In addition a two-dimensional generalisation of the DACP was constructed~\cite{Blythe2001} and
used to study wetting and interface phenomena~\cite{BlytheEvans2001}.

\section{Numerical evidence for a discontinuous velocity transition}

We begin our study of the DACP by presenting numerical data that suggests the velocity of the wave emanating from the left boundary is discontinuous at the phase transition point.

\subsection{Simulation details}

We performed direct Monte Carlo simulations of the microscopic DACP dynamics 
specified in section~2 above. Each run was initialised with the lattice unoccupied,
apart from the 0$^{\rm th}$ site which is kept permanently active.  During the
simulation a list of active sites is maintained.  In each elementary update,
one of the $n$ active sites is chosen from this list at random.  With probability
$\frac{r}{r+1}$, its right neighbour is activated; otherwise the chosen active site becomes
inactive.

Since our aim is to measure a velocity, it is important to keep track of the length of
time associated with each update.  Since we attempt an update on any of the $n$ active sites
the total attempt rate is $\lambda = n(1+r)$.  In principle, the size of the time
increment, $\Delta t$, for each step should be sampled from an
exponential distribution with mean $1/\lambda$. However since the time
increments of any realisation are small the distribution becomes
sharply peaked and we can instead use the mean of the distribution as
the size of our timestep. This approximation is computationally
favourable and we have verified that it made no difference to the
results. Thus we take the time taken for each update to be $\Delta t =
\frac{1}{n(1+r)}$, where $n$ is the total number of active sites
before the update takes place.

The front velocity of a single run was defined as the position of the
rightmost active site at the end of the run, divided by the time
taken for the run to end. This definition was chosen since it is
unambiguous in the microscopic model.  The simulations were run until a
prespecified amount of time had elapsed: the lattice expanded as needed to
accommodate all active sites. To reduce the noise
in the data the simulation was repeated for 500 runs for each value of
$r$ and the quantities of interested averaged over this ensemble of runs.

\subsection{Results}

\begin{figure}[tb]                     
        \begin{center}
                \leavevmode             
                \epsfclipon
                \epsfxsize=0.5\linewidth         
                \epsffile{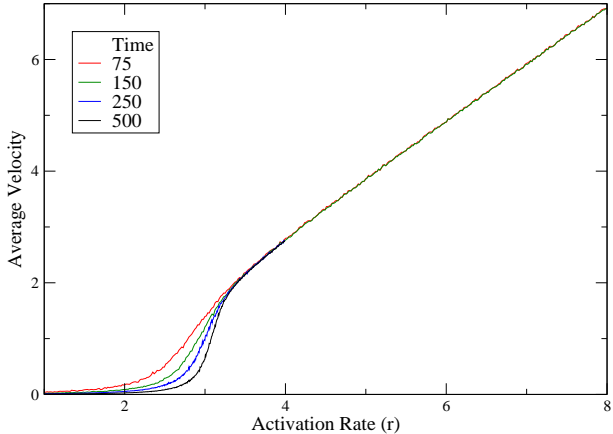}
        \end{center}
    \caption{A plot of the velocity vs. activation rate for different run times.
The two shorter
runs spanned activation rates in the interval $r \in [1,8]$ and the
longer two $r \in [1,4]$, in both cases the intervals were subdivided
into 500 equally spaced values of $r$ for which the simulation was
run.
Note that as the simulation time is increased a discontinuity appears to develop.}
    \label{fig:DiscontinuityTReal}
\end{figure}

The first interesting result pertains to the asymptotic front velocity.
The simulation was run for several different simulation times as shown in \figref{fig:DiscontinuityTReal}.
Our definition of the front velocity implicitly assumes
that an asymptotically constant velocity is reached on a timescale short compared to
that at which the simulation ends.  We anticipate that this may not be the case for
values of $r$ near the critical value $r_c$, since relaxation times diverge here.
Therefore measurements of the velocity near $r_c$ are subject to finite-time corrections.
As can be seen from \figref{fig:DiscontinuityTReal}, increased run time shows a sharpening of
the velocity as a function of $r$ near $r_c$, suggestive of a discontinuous transition in the
infinite-time limit.

\begin{figure}[tb]
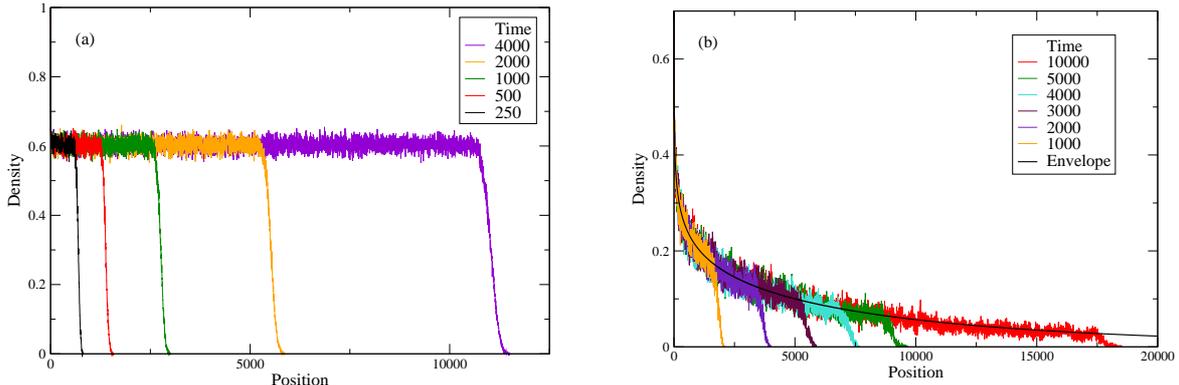

 \centering
 \begin{minipage}[c]{0.45\linewidth}
    \begin{center}
        \leavevmode             
        \epsfclipon
        \epsfxsize=\linewidth        
        \epsffile{SupercriticalFor1stYear.eps}
    \end{center}
 \end{minipage}
 \hspace{0.05\linewidth}
 \begin{minipage}[c]{0.45\linewidth}
    \begin{center}
        \leavevmode             
        \epsfclipon
        \epsfxsize=\linewidth        
        \epsffile{EnvelopeFor1stYear.eps}
    \end{center}
 \end{minipage}
\caption{(a) Density vs.\ position in the supercritical regime at different times. A wave of constant bulk density invades the empty lattice as time progresses. (b) Corresponding data for the subcritical regime. The wave is attenuated by an envelope fitted empirically by the solid line.}
\label{fig:TwoImages}
\end{figure}

We also studied the density profile in the two different phases.  The position and shape of the front
can be investigated by looking at the density
profile for a single activation rate at different times. Choosing two
values of $r$ close to, and on either side of, the critical rate we
observe two very different scenarios. For the supercritical case
(\figref{fig:TwoImages}a) there is a wave of constant bulk density
invading the empty lattice. The front simply propagates away from
the boundary at a constant velocity.

For the subcritical case (\figref{fig:TwoImages}b) we observe 
an attenuated wave: the front propagates away from the boundary whilst at the same time decaying away. However the decaying density seems to follow an envelope of sorts. As a guide to the eye this envelope has been added to the graph.  We will return to the idea of this envelope again in section~\ref{sec:F-KPP-theory}.

It is perhaps surprising at a first glance that the \emph{continuous} DP transition should be accompanied by a \emph{discontinuity} in the front velocity. In the remainder of this work, we
elucidate the mechanism behind this discontinuity.

\section{Directed Percolation scaling picture}

Despite the absence of an absorbing state in the DACP, we show in this section that the behaviour of the DACP
described in the previous section can in fact be interpreted within the universal scaling picture associated with the Directed Percolation phase transition.

\subsection{Steady-state density profile}

We shall consider the steady state density profile of the system
$\rho_j = \lim_{t\to\infty} \langle \tau_j \rangle$.
A suitable order parameter  for the system is the steady-state
density  as $j\to \infty$, which we denote by $\overline{\rho}$.
In the supercritical phase, close to criticality, we expect the order parameter
to scale as $\overline{\rho} \sim \Delta^\beta$, where $\Delta = \abs{r-r_c}$ is the distance from
criticality and $\beta$ is the order parameter exponent which we expect
to be equal to the value for DP in one spatial dimension.
We can also define a characteristic length scale $\xi$  as
the steady-state density decay length 
in the subcritical phase where $\rho_j \sim {\rm e}^{-j/\xi}$. Then, since the length scale
diverges at criticality we expect it to scale as $\xi \approx
\Delta^{-\nu}$ near criticality where $\nu$ is a correlation length exponent. 
At criticality we expect the  profile to decay as a power law with exponent $\delta$.

Near criticality we further expect the scaling form $\rho_j \sim j^{-\delta}g(j/\xi)$
with $g(u)$ a scaling function obeying $\lim_{\xi\rightarrow\infty} g(j/\xi) =
g(0) = \textrm{constant}$ such that at criticality $\rho_j \sim
j^{-\delta}$. However since the steady-state density approaches a
non-zero constant as $x\rightarrow\infty$ in the supercritical phase
it follows that $\lim_{u\rightarrow\infty} g(u) \sim u^\delta$. Thus
$\lim_{j\rightarrow\infty} \rho_j \sim j^{-\delta} u^\delta =
\xi^{-\delta} \sim \Delta^{\delta\nu}$. But we also know that
$\lim_{j\rightarrow\infty} \rho_j  = \overline{\rho} \sim \Delta^\beta$. Thus it
follows that $\delta=\frac{\beta}{\nu}$.

\begin{figure}[tb]                     
        \begin{center}
                \leavevmode             
                \epsfclipon
                \epsfxsize=0.5\linewidth         
                \epsffile{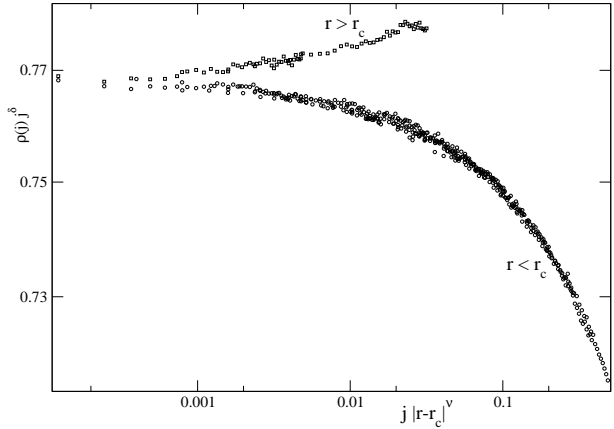}
        \end{center}
    \caption{Collapse of the steady-state density profile $\rho_j$ for a range of activation rates $r$ onto two master curves, 
    one for the supercritical regime (upper curve) and one for the subcritical regime (lower curve). The collapse was obtained
    by varying the free parameters $r_c$, $\delta$ and $\nu$.  Figure reproduced from \cite{Blythe2001}.}
    \label{fig:Collapse}
\end{figure}

In \cite{Blythe2001}, the critical point $r_c$ and exponents $\delta$ and $\nu$ were obtained
by plotting $\rho_j j^\delta$ against $u=j/\xi = j |r-r_c|^{\nu}$ for different values of $r$,
and varying $r_c$, $\delta$ and $\nu$ until the best data collapse (as judged by eye) was obtained.
The resulting collapse, obtained for $r_c=3.3055(5), \delta=0.1640(5)$ and $\nu=1.7(2)$, is reproduced
from \cite{Blythe2001} in \figref{fig:Collapse}.

We compare these measurements with the established values of the DP exponents in table~\ref{tab:DPexponents}.  We observe good agreement with the DP values that apply in one spatial dimension (1D) as long as we identify the exponent $\nu$ with the DP \emph{temporal} correlation exponent $\nu_\parallel$, as opposed to the distinct, and independent, spatial exponent $\nu_\perp$: we discuss this point in more detail shortly.  For future reference, we have also included the exponents obtained within a mean-field approximation in table~\ref{tab:DPexponents}. For an in-depth study of DP exponents and scaling
we refer the reader to~\cite{Hinrichsen2000}. Here, we focus more on the dynamic behaviour of the DACP and the behaviour of the active front as it moves out from the left boundary.

\begin{table}[h!t!b!]
  \begin{center}
    \begin{tabular}{l | c c r}
      & DP (MF)~\cite{Hinrichsen2000} & DP (1D)~\cite{Jensen1999} & DACP~\cite{Blythe2001}\\ \hline
      $\delta$ & 1 & 0.159464(6)& 0.1640(5)\\
      $\nu_\parallel$ & 1 & 1.733847(6) & 1.7(2) \\
      $\nu_\perp$ & 1/2 & 1.096854(4)&  \\
    \end{tabular}
  \end{center}
  \caption{Critical exponents for Directed Percolation (DP) in a mean field approximation (MF) and one spatial dimension (1D), and for the Driven Asymmetric Contact Process (DACP) of the present work.}
  \label{tab:DPexponents}
\end{table}

\subsection{Advection dynamics: the shearing of DACP into DP}
\label{sec:DACP2DP}

\begin{figure}[tb]                     
        \begin{center}
                \leavevmode             
		\epsfclipon
                \epsfysize=50mm         
                \epsffile{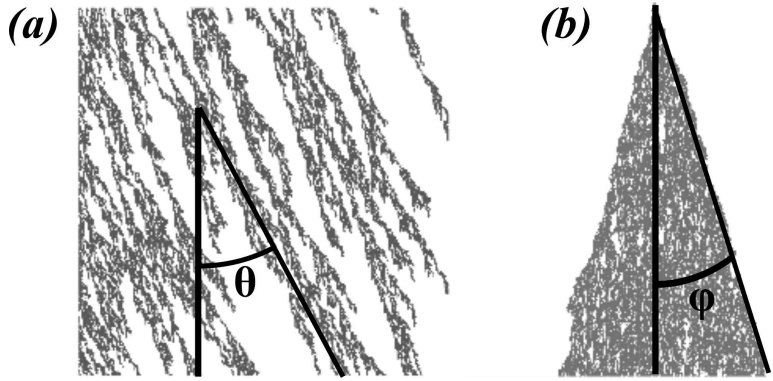}
        \end{center}
    \caption{Space time plots of (a) the DACP and (b) DP.  In DP, a cone of activity spreads at an angle $\phi$ to the time (vertical) axis; to obtain the DACP, this cone is sheared by a further angle $\theta$ due to the advection.  Note that the activation rates have been chosen to allow easy identification of the two angles in these figures: the distance from the critical point is different in the two cases, and hence there is no significance of the density of the DP clusters being greater than that for the DACP.}
    \label{fig:DACPandDP}
\end{figure}

The easiest way to understand the effect of asymmetry (or advection)
in the contact process is to directly compare a space-time plot of the
DACP dynamics with its contact process counterpart
and its description in terms of DP scaling exponents---see
\figref{fig:DACPandDP}.  From these plots we identify two
characteristic angles.  In the DACP, \figref{fig:DACPandDP}a, the
activity emanates from the wall with an axis that is an angle $\theta$
to the vertical.  This axis corresponds to the time direction in the
basic contact process, \figref{fig:DACPandDP}b. The leading edge of
the activity emerges at an angle $\phi$ to the DP time axis.  We thus
picture the DACP as a spatial shearing of DP by the shear element $\tan\theta$.
One consequence of this shearing is that the spatial correlation
length measured in the steady-state density profile of the DACP is a
linear combination of the temporal and spatial correlations lengths in
DP. Such that $\xi_\perp' = \xi_\perp + \xi_\parallel\tan\theta$ and $\xi_\parallel' = \xi_\parallel$, with primed variables pertaining to the DACP.  Hence, we expect the DACP correlation length to diverge with the
faster-growing DP correlation length as $r\to r_c$, and therefore that
$\nu$ should be identified with the larger of the DP exponents (i.e.,
$\nu_\parallel$, as above).

We now turn to the dynamics.  Above the critical point in DP, an activity wave travels at a velocity $v_{\rm DP} = \tan \phi$.  If, near criticality, there is a single characteristic length and time scale, this velocity must be given by their ratio. That is
\eqn{\label{eqn:vDP}
  v_{\rm DP} = \tan\phi = \frac{\xi_\perp}{\xi_\parallel} \sim \Delta^\chi \;\;\;\mbox{with}\;\;\; \chi = {\nu_\parallel-\nu_\perp} \;.
}
Since $\nu_{\parallel} > \nu_{\perp}$, this velocity vanishes as $r \to r_c$ from above.  Below criticality, activity does not spread out indefinitely, and so the wave velocity is zero when $r<r_c$.  Thus, the wave velocity is continuous across the transition in the absence of advection.

We obtain the supercritical wave velocity for the DACP, $v_{\rm sup}$, by applying the shear $\tan\theta$.  That is,
\eqn{\label{eqn:VelocityExponent}
  v_{\rm sup} = \frac{\xi_\perp'}{\xi_\parallel'} = \frac{\sin\phi +\tan\theta\cos\phi}{\cos\phi} = \tan\phi + \tan\theta = v_{\rm DP}+s = s + A\Delta^\chi
}
where we have introduced the \emph{advection} $s = \tan\theta$.  (There may be subleading corrections coming from
corrections to scaling in the relation (\ref{eqn:vDP}).)  The actual value
of $s$ is a nontrivial emergent property of the stochastic DACP
dynamics.  However, a key point is that it can take a nonzero value in
the subcritical regime which yields the {\em intrinsic} velocity $v_{\rm sub}$: 
activity still propagates at 
rate $v_{\rm sub} =s$ from the active boundary, despite the fact
that it dies out after some finite time.  Thus, the {\em apparent} 
(i.e. the observed)  wave velocity, measured far from the origin,  
may jump discontinuously from zero to some nonzero value
$v_c=s(r_c)$ at the critical point. This is in contrast to the DP wave velocity which goes to zero at criticality, as shown above.  
The leading-order behaviour of the supercritical wave
velocity in the DACP is then $v_{\rm sup} \sim v_c + A \Delta^\chi$
where $\chi=\nu_\parallel - \nu_\perp$.  We remark that this
expression provides a means to measure the smaller correlation length
exponent, $\nu_\perp$, which cannot be accessed from the steady-state
density profile alone.

\section{Mean-field theory}
\label{sec:F-KPP-theory}

Having proposed a mechanism for the discontinuous velocity transition observed in section~3, we now demonstrate explicitly that it is at work within the mean-field formulation of the DACP dynamics.

We know that for a DP model, such as the contact process, the
continuous-space mean-field description is given by a Fisher-KPP
equation. The Fisher-KPP equation was first introduced in~\cite{Fisher1937,
Kolmogorov1937, Luther1906}, see~\cite{Saarloos2003} for a
review.  We begin with the general form of the Fisher-KPP equation
supplemented with an advection term to take into account the boundary
drive present in the DACP:
\eqn{\label{eqn:FKPPa} 
\frac{\partial\rho}{\partial t} = \alpha\rho -
\beta\rho^2 - r'\frac{\partial \rho}{\partial x} + D\frac{\partial^2
\rho}{\partial x^2}
 } 
At this stage  $\alpha, \beta, D$ and $r'$ are phenomenological parameters.  We will
match these up with the parameters of the microscopic model in section~\ref{sec:params} below.
Meanwhile we note that we expect $r'$ and $D$ to be smooth functions of
the microscopic advection coefficient $r$ and that $\alpha \propto r-r_c$.

We first establish the two distinct stationary regimes
which are determined by the sign of $\alpha$.  If we set
$\frac{\partial\rho}{\partial t} =0$ in~\eqnref{eqn:FKPPa} we find
the stationary density $\rho^*(x)$ is given by
\eqn{\label{eqn:Stationary} 0 = \alpha\rho^* - \beta{\rho^*}^2 -
r'\frac{\partial \rho^*}{\partial x} + D\frac{\partial^2
\rho^*}{\partial x^2}\;. 
} 
The solution takes the following forms for large $x$.

\begin{description}
\item[For $\alpha <0$ ({\em subcritical case)}]:   a density profile which decays exponentially to zero
and  has the form for large $x$,  $\;\rho_{\rm sub}^*(x)\simeq A e^{-\lambda x}$ 
\item[For $\alpha >0$ ({\em supercritical case})]: 
a density profile which decays exponentially to a non-zero value and has the form for large $x$,
$\;\rho_{\rm sup}^*(x) \simeq \overline{\rho} + Be^{-\lambda x}$ with $\overline{\rho} =\frac{\alpha}{\beta}$.
\end{description}

In both cases the decay constant may be written as
\eqn{ \lambda =    \frac{-r' + \left(r'^2+ 4 |\alpha|D\right)^{1/2}}{2D} .
}
For small $|\alpha|$, $\lambda= O(|\alpha|)$.  Thus the characteristic lengthscale $\xi$ diverges as $|\alpha|^{-1}$
at criticality.  Following the arguments of section 4, $\xi$ is expected to diverge with the DP exponent $\nu_{\parallel}$.  From table~\ref{tab:DPexponents}, we see indeed that $\nu_{\parallel}=1$ in the mean field.

Based on the envelope observed in the simulation of the stochastic system (see section 3) we assume that the full time-dependent
density can be described by $\rho(x,t) = \rho^*(x)f(x,t)$ where $\rho^*(x)$ is the stationary solution to~\eqnref{eqn:Stationary}. Equation (\ref{eqn:FKPPa}) then becomes
\eqn{
  \rho^*\dot{f} = \alpha{\rho^*} f - \beta {\rho^*}^2 f^2 - r'({\rho^*}'f + {\rho^*} f') +D({\rho^*}''f +2{\rho^*}'f' + {\rho^*} f'') \;.
} 
Dividing through by ${\rho^*}$ and using the definition of ${\rho^*}$, \eqnref{eqn:Stationary}, to eliminate $\alpha$ we obtain
a modified Fisher-KPP equation for the wave $f(x,t)$ that sits inside the envelope:
\eqn{\label{eqn:f-equation}
  \frac{\partial f}{\partial t} = \beta\rho^*f(1-f) - s\frac{\partial f}{\partial x} + D\frac{\partial^2 f}{\partial x^2} \;.
}
In this equation, the advective velocity of the modified wave, $s$, is given by
\eqn{\label{eqn:sdef}
 s = r' - \frac{2D{\rho^*}'}{\rho^*} \;.
}
Note that this advective velocity, $s$, is generally increased over the bare quantity $r'$ as a consequence of the envelope.
The coefficient, $\beta \rho^*(x)$ of the non linear growth term $f(1-f)$ is generally $x$ dependent, but in the
supercritical phase it decreases to a constant  value for large $x$ thus recovering the Fisher-KPP equation far from the boundary. On the other hand, in the subcritical regime the non-linear term decreases to zero far from the boundary.

First we consider the supercritical phase.
Far from the boundary (\ref{eqn:f-equation}) becomes, using $\rho^* \to \alpha/\beta$
and $s \to r'$
\eqn{
  \frac{\partial f}{\partial t} = \alpha f(1-f) - r'\frac{\partial f}{\partial x} + D\frac{\partial^2 f}{\partial x^2} 
}
which is  the usual Fisher wave equation with advective coefficient $r'$.
Following the usual approach (see e.g. \cite{Saarloos2003}), one assumes a  travelling wave form
$f(x,t)=f(x-vt)=f(z)$ and  linearises for large $z$ where $f$ is small. The solution is of  the form $f=e^{-\mu z}$ with
\eqn{
  \mu = \frac{(v-r') \pm \sqrt{(v-r')^2 - 4D\alpha}}{2D} \;.
} 
Since $\mu \in \mathbb{R}_+$ we require $(v-r')^2 \geq 4D\alpha \Rightarrow v \geq r' + \sqrt{4D\alpha} = v_{\min}$.
We now assume (as for the normal Fisher wave starting from an initial sharp front) that $v=v_{\min}$. Thus
\eqn{\label{eqn:vsup}
v_{\rm sup} = r' + \sqrt{4D\alpha}\;.
}
Of course (\ref{eqn:vsup}) could simply be obtained by adding the advective velocity $r'$ to the usual Fisher
wave velocity in the absence of advection, $\left(4\alpha D\right)^{1/2}$. 

In the subcritical phase, the modified wave 
described by (\ref{eqn:f-equation})
has a velocity given by a distinct expression.  Here, for large $x$, \eqnref{eqn:f-equation} becomes
\eqn{\label{eqn:LongTimeSubCrit}
  \dot f = A \beta e^{-\lambda x}f(1-f) - sf' + Df'' 
}
where
\eqn{\label{eqn:s1}
  s= r' + 2 D \lambda = (r'^2 + 4 \abs{\alpha} D)^{1/2}\;. 
}
Note that the advective velocity of the attenuated wave is increased over the value $r'$.

The spatial dependence of the coefficient of the non-linear term makes the analysis of equation \eqnref{eqn:LongTimeSubCrit} non-trivial.
Here we content ourselves with  a heuristic picture.
Initially the presence of the nonlinear term will 
mean that a nonlinear travelling wave emanates from the boundary.
However as the front moves away from the boundary
the nonlinear term becomes less important  and we expect the front to broaden and the velocity to decrease.
Finally as the front of the wave moves further away and  $x \gg 1/\lambda$
the equation for $f $ reduces to a diffusion equation with advection
\eqn{\label{eqn:diffront}
\dot f = - sf' + Df'' \;.
}
The wavefront thus broadens diffusively over time.
Therefore, at late times, the modified density profile $f$ will be approximately that given by a diffusion equation.
This diffusive front  moves with velocity
\eqn{\label{eqn:vsub}
  v_{\rm sub} = s = (r'^2 + 4 |\alpha| D)^{1/2}\;,
}
and the profile itself takes the form
\eqn{
  f(x,t) \simeq  \frac{1}{2} \erfc\left[\frac{x-st}{2\sqrt{Dt}}\right] \;.
}
Thus the width of the front is ultimately $\sqrt{Dt}$.

To summarise, the analysis of the phenomenological Fisher-KPP equation
\eqnref{eqn:FKPPa} shows that there are two possible regimes, according to the sign of $\alpha$.
When $\alpha$ is negative, the density decays to zero as it moves away from the origin; when positive,
it propagates away from the origin with a constant bulk density at a constant velocity.  The dynamics can
be couched in terms of a modified wave, $f(x-vt)$, travelling within the envelope of the steady-state density profile, $\rho^*(x)$.
This wave is governed by the modified Fisher-KPP equation, \eqnref{eqn:f-equation}, and has a velocity in these
two states of $v_{\rm sub} = \sqrt{r'^2 -4D\alpha}$ and $v_{\rm sup} = r' + \sqrt{4D\alpha}$. 
Note that $v$ is continuous at the transition $\alpha=0$ although its derivative is not.

In the subcritical regime, this velocity coincides with the advective velocity $s$ introduced
in section 4 in terms of the angle of shear from DP to the DACP.
As noted previously, one effect of the envelope is to force the apparent velocity (i.e., that
observed in simulations) to decay to zero: activity can probe only a finite distance from the origin as
time $t\to\infty$.  Above the critical point, the observed velocity is $v_{\rm sup}$ and so, across the
transition, this observed velocity exhibits a discontinuity.  We note from the form of $v_{\rm sup}$, the
velocity grows from its critical value as $v-v_c \sim \alpha^\chi$ where $\chi= 1/2$.  This value of $\chi$
agrees with the prediction $\chi = \nu_{\parallel} - \nu_{\perp}$ from the mean-field theory for DP (see table~\ref{tab:DPexponents}).

\section{Modified travelling wave in the stochastic DACP}

We now revisit the simulations of the DACP in the light of what we have learnt from the mean-field
theory.  We first express the phenomenological parameters appearing in the Fisher-KPP equation, \eqnref{eqn:FKPPa},
in terms of the stochastic activation rate $r$.  We then examine more closely the numerical evidence
for the picture of an modified wave travelling within the stationary density profile, and in particular,
the prediction for the growth of the wave velocity just above criticality, i.e., $v-v_c \sim \alpha^\chi$ where
$\chi = \nu_{\parallel} - \nu_{\perp}$ given by the appropriate DP exponents.

\subsection{Identification of the mean-field phenomenological parameters}
\label{sec:params}

We first revisit the mean-field theory of Section~\ref{sec:F-KPP-theory}
and compare to the microscopic DACP dynamics.
One may make and make a heuristic identification of the phenomenological parameters in (\ref{eqn:FKPPa})
by considering first of all an exact equation for the rate of change of density at site $i$ in the DACP.
This reads
\eqn{
  \deriv{}{t}\av{\tau_i(t)} = r\av{\tau_{i-1}(t)\left[1-\tau_i(t)\right]}-\av{\tau_i(t)} \;,
}
where the first term comes from site $i-1$ activating site $i$ at rate $r$ if the former is active and the latter inactive,
and the second term from the decay of site $i$ at unit rate when it is active.  The mean-field approximation
is to write $\langle \tau_i(t) \tau_j(t) \rangle = \rho_i(t) \rho_j(t)$, where $\rho_i(t) = \langle \tau_i \rangle$.
\eqns{\label{eqn:mf}
  \dot{\rho_i} &= r\rho_{i-1}(1-\rho_i)-\rho_i \;\;\; \forall i \geq 1 \\
  \rho_0 &= 1 \;\;\; \forall t
}
The steady state solution of this spatially discrete equation is provided in the appendix.

Here we move over to continuous space and expanding to second order
spatial derivatives, we find, $\alpha=r-r_c$, $\beta=r$, $r'=r(1-\rho)$
and $D=\frac{r'}{2}$, where $r_c = 1$. This suggests a density dependence in $r'$ and
$D$: however, as we now explain, this is not expected to affect the
wavefront behaviour.

In the subcritical regime the density profile tends to zero far from the boundary therefore $r'\to r$ and $D\to r/2$.
Thus 
\eqn{\label{eqn:vsubstoch}
  v_{\rm sub} = \sqrt{r^2 -2r\alpha} = \sqrt{r_c^2 -\alpha^2} \simeq r_c  - \frac{\alpha^2}{2 r_c}  +\ldots
}

In the supercritical regime, on the other hand, $\rho$ will tend to zero at the tip of the wavefront but  be nonzero behind the Fisher wavefront. Thus at the tip $r'= r$ and the supercritical Fisher wave velocity becomes
\eqn{\label{eqn:vsupstoch}
  v_{\rm sup} = r + \sqrt{2r\alpha} \simeq  r_c + \sqrt{2r_c\alpha}\;.
}
Therefore the derivative of the modified wave velocity is discontinuous at the phase transition. It is also interesting to note that with our identification of $r'$ and $\beta$, behind
the wavefront in the supercritical regime  we have $r'=r(1-\alpha/r) = r_c$ and the effective advection is fixed at the critical value.

\subsection{Stochastic modified wave dynamics}

Returning now to the stochastic simulations of the DACP, we investigate first the picture of a modified wave travelling within a
density envelope.  In \figref{fig:TwoImages}b, the density profile in the subcritical phase is shown at different time points along with an envelope of the form $\rho^\ast(x) = \exp(-A_0x-A_1x^{A_2})$, with the fitting parameters $A_i$ all positive.  As we get closer to criticality $A_2$ decreases, indicating an approach to the scaling form $e^{-x/\xi}x^{-\delta}$.

\begin{figure}[tb]
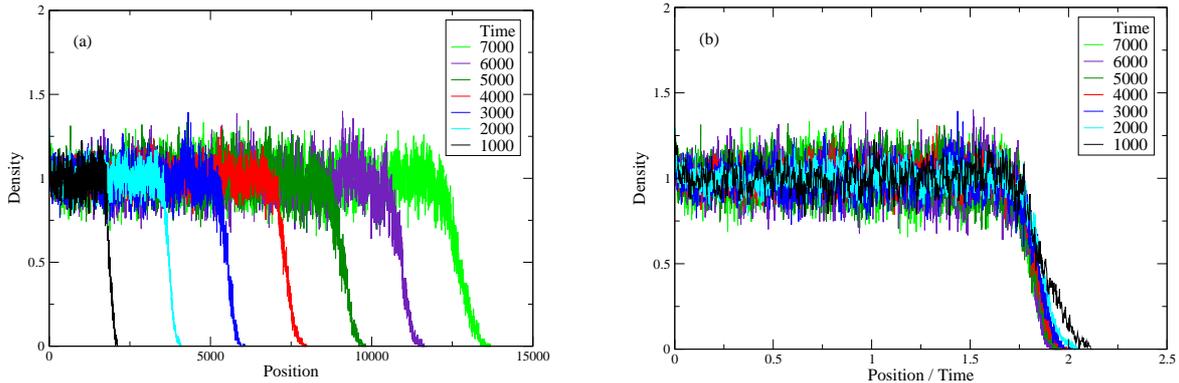

 \centering
 \begin{minipage}[c]{0.45\linewidth}
    \begin{center}
        \leavevmode             
        \epsfclipon
        \epsfxsize=\linewidth        
        \epsffile{DensityPos,SubCritical,4,again3.3.eps}
    \end{center}
 \end{minipage}
 \hspace{0.05\linewidth}
 \begin{minipage}[c]{0.45\linewidth}
    \begin{center}
        \leavevmode             
        \epsfclipon
        \epsfxsize=\linewidth        
        \epsffile{DensityPos,SubCritical,5,again3.3.eps}
    \end{center}
 \end{minipage}
\caption{(a) The density vs. position with the envelope divided out ($r=3.3$). (b) The density vs. position/time. The collapse shows that the velocity remains constant.}
\label{fig:EnvelopelessAndCollapse}
\end{figure}

To obtain the modified wave $f(x-vt)$, we divide the numerical density profiles by this envelope equation.  The result of this procedure is shown in \figref{fig:EnvelopelessAndCollapse}a, which clearly shows a wave with constant bulk density invading an empty lattice.  The constancy of the wave velocity can be checked by dividing the $x$ coordinate by time (\figref{fig:EnvelopelessAndCollapse}b).  We do not, however, see strong evidence for the diffusive broadening of the wavefront predicted by \eqnref{eqn:diffront}. This could be because this equation applies only where the stationary density is small, a region that is a hard to access numerically.

\begin{figure}[tb]                     
        \begin{center}
                \leavevmode             
                \epsfclipon
                \epsfxsize=0.5\linewidth         
                \epsffile{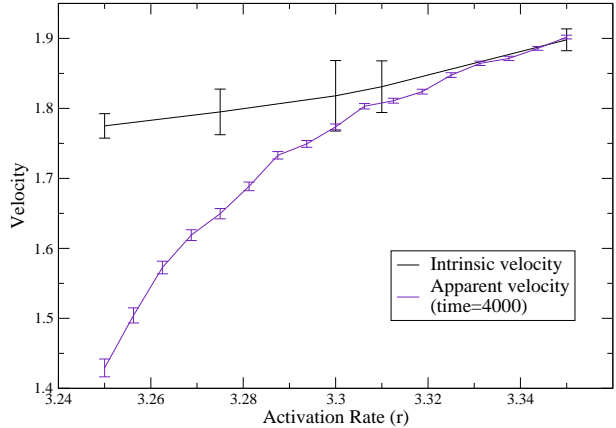}
        \end{center}
    \caption{The intrinsic velocity: Velocity vs. activation rate. We see a distinct difference between the apparent and intrinsic velocities. Note that the apparent velocities only seems continuous due to finite-time effects.}
    \label{fig:IntrinsicVelocity}
\end{figure}

A key component of the scaling picture (section 4), seen explicitly
within the mean-field dynamics (section 5), is the continuity of the
intrinsic  wave velocity across the transition.  After
unfolding the envelope (as in \figref{fig:EnvelopelessAndCollapse})
one can obtain this velocity over a range of activation rates $r$.  In
\figref{fig:IntrinsicVelocity}, this velocity is compared with the
apparent velocity before unfolding the velocity (i.e., the data of
\figref{fig:DiscontinuityTReal}).  The two velocities are clearly
distinct in the subcritical regime ($r<r_c \approx 3.3$). Since the
envelope decays exponentially in the subcritical phase, it is
difficult to probe the late-time travelling-wave dynamics, and so the
error bars on the modified wave velocity are necessarily large.
However, the data suggest that the intrinsic velocity changes more slowly just
below the critical point than above, in qualitative agreement with the
mean-field predictions of equations \eqnref{eqn:vsubstoch},\eqnref{eqn:vsupstoch}.

\begin{figure}[tb]                     
        \begin{center}
                \leavevmode             
                \epsfclipon
                \epsfxsize=0.5\linewidth         
                \epsffile{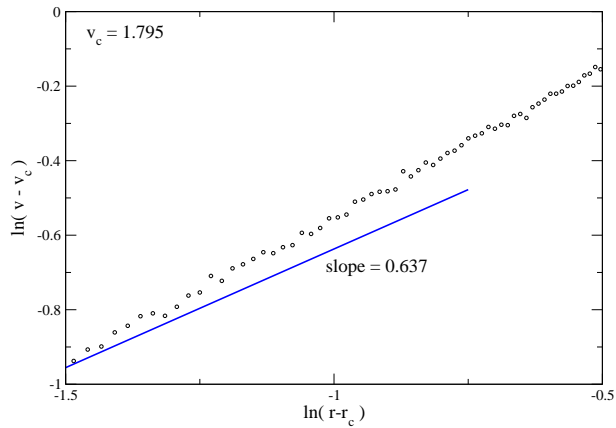}
        \end{center}
    \caption{Velocity exponent $\chi$ measured in the stochastic simulations.  Shown is $\ln (v - v_c)$ against $\ln (r-r_c)$ with $v_c=1.795$ to obtain a straight line in the regime $r\to r_c$.  The solid line has the gradient $\chi \approx 0.637$ appropriate for Directed Percolation in one dimension.}
    \label{fig:velexponent}
\end{figure}

Finally, we attempt to access the exponent $\chi =
\nu_{\parallel}-\nu_{\perp}$ by fitting the observed wave velocity in
the supercritical regime to the form $v_{\rm sup} \sim v_c + A
(r-r_c)^\chi$ suggested by the scaling picture of section 4 and
confirmed within the mean-field regime by travelling wave analysis of
section 5. It turns out that an estimate of $\chi$ is rather sensitive
to the values of $v_c$ and $r_c$ used in a straight-line fit to $\ln
(v-v_c)$ plotted as a function of $\ln (r-r_c)$.  Taking $r_c=3.3055$,
we find the best fit (as quantified by the sum of square residuals)
when $v_c\approx1.795$. The corresponding plot is shown in
\figref{fig:velexponent} along with a line of gradient $\chi=0.637$,
which is the appropriate choice for DP in one dimension.  We see that
the simulation data have a gradient that is consistently higher than
the predicted value.  It may be that corrections to the leading-order
behaviour remain significant in the region we have been able to access
numerically.  We remark that reasonable straight lines are obtained
for values of $v_c$ in the range $1.70$ to $1.84$, yielding estimates
of $\chi$ between $0.6$ and $0.8$ suggesting that the numerical data
are consistent with our scaling prediction for $\chi$.

\section{Conclusion}

In this work we have studied a variation of the contact process which
includes an active boundary which drives the system and advection away
from that boundary. As expected the system exhibits a phase transition
from a state with activity localised near the boundary to a state
where a wave of activity emanates away from the boundary. We identify
the DP critical exponents $\beta$ and $\nu_\parallel$ by considering
the behaviour of the density far from the boundary and the spatial
decay length over which the density decays to that value.  On the
other hand, the velocity of the activity wave emanating from the
boundary exhibits some perhaps unexpected behaviour: in the
subcritical phase the apparent velocity is zero whereas in the
supercritical phase the velocity jumps discontinuously to a non-zero
value.  We have explained this phenomenon by studying a mean field
theory in which we show that an intrinsic velocity for a wave
emanating from the boundary  and described by
(\ref{eqn:f-equation}),
exists both below and above the
transition.  However in the subcritical regime the spatially decaying
envelope for this wave means that the apparent velocity observed in
the simulations is zero.  This picture appears to hold well in
simulations of the stochastic system

The study raises several interesting questions. 
At the mean field level  it would be
interesting to put our analysis of the modified Fisher equation
(\ref{eqn:FKPPa}), in particular the subcritical case, on a more rigorous
footing.  A study of the possible crossover in the solution of
(\ref{eqn:FKPPa}) from a nonlinear wave to a diffusive wave would be
illuminating.

The noisy version of the Fisher-KPP  equation is known to describe 
contact processes  \cite{Muller1995}. It would of course be of great interest
to further understand travelling wave solutions 
of the noisy  version of the Fisher-KPP equation,
in particular the velocity and width of the front.

\appendix

\section{Appendix: Solution of spatially discretized mean field equation}

The mean-field equation (\ref{eqn:mf}) governing the density in the DACP process is
\eqns{\label{eqn:mf2}
  \dot{\rho_i} &= r\rho_{i-1}(1-\rho_i)-\rho_i \;\;\; \forall i \geq 1 \\
  \rho_0 &= 1 \;\;\; \forall t
}
The steady-state solution ($\dot{\rho_i} =0$) is
\eqn{ \rho_i = \frac{(1-r)r^i}{1-r^{i+1}}}
which yields the large $i$ behaviour
\begin{eqnarray}
 \rho_i &\simeq&  (1-r)r^i\qquad\qquad\qquad\;\;\mbox{for}\quad r<1 ,\\
 \rho_i &=&  \frac{1}{i+1}\qquad\qquad\qquad \qquad \mbox{for}\quad r=1 ,\\
 \rho_i &\simeq&  \frac{(r-1)}{r}\left(1+ r^{-(i+1)}\right)\quad\mbox{for}\quad r>1 . 
\end{eqnarray}
Thus this mean field theory is consistent with 
the profiles decaying a decay length  which diverges as  $1/|\ln r| \sim \Delta^{-1}$
and the order parameter emerging as $ \overline{\rho} \sim \Delta$
where $\Delta = r-r_c$ with $r_c =1$.
Also at criticality we have a power law decay of the profile with exponent
$\delta =1$. All these exponents are consistent with the mean field DP exponents given in Table~~\ref{tab:DPexponents}.

%
\bibliographystyle{unsrt}                      

\addcontentsline{toc}{section}{References}{}
\bibliography{biblio}       

\end{document}